\documentclass[a4paper,12pt]{article}

\usepackage{amsmath}
\usepackage{amssymb}

\makeatletter
\@addtoreset{equation}{section}
\renewcommand{\theequation}{\thesection.\@arabic\c@equation}
\makeatother

\makeatletter
\renewcommand\appendix{\par
  \setcounter{section}{0}%
  \setcounter{subsection}{0}%
  \gdef\thesection{Appendix \@Alph\c@section }
  \renewcommand{\theequation}
  {\Alph{section}.\arabic{equation}}
}
\makeatother

\newcommand{\Dhat}{\hat{D}}
\newcommand{\Sbb}{\mathbb{S}}
\newcommand{\gbar}{\bar{g}}

\newcommand{\ebar}{\bar{\varepsilon}}
\newcommand{\pbar}{\bar{P}}

\newcommand{\etahat}{\hat{\eta}}
\newcommand{\zetahat}{\hat{\zeta}}

\begin{document}

\begin{titlepage}

\vspace*{-15mm}   
\baselineskip 10pt   
\begin{flushright}   
\begin{tabular}{r}    
{\tt KEK-TH-1693}\\   
December 2013
\end{tabular}   
\end{flushright}   
\baselineskip 24pt   
\vglue 10mm   

\begin{center}
{\Large\bf
 Compressible Fluids in the Membrane Paradigm: 
 non-AdS Fluid/Gravity correspondences
}

\vspace{8mm}   

\baselineskip 18pt   

\renewcommand{\thefootnote}{\fnsymbol{footnote}}
\setcounter{footnote}{1}

Yoshinori~Matsuo$^{a}$\footnote{E-mail: ymatsuo@post.kek.jp}
and 
Yuya~Sasai$^b$\footnote{E-mail: sasai@law.meijigakuin.ac.jp}

\renewcommand{\thefootnote}{\arabic{footnote}}
\setcounter{footnote}{0}
 
\vspace{5mm}   

{\it  
$^a$ KEK Theory Center, High Energy Accelerator Research Organization (KEK), Tsukuba 305-0801, Japan \\
$^b$ Institute of Physics, Meiji Gakuin University, Yokohama 244-8539, Japan  
}
  
\vspace{10mm}   

\end{center}

\begin{abstract}
Correspondences between  black holes and fluids 
have been discussed in two different frameworks, 
the Fluid/Gravity correspondence and membrane paradigm. 
Recently, it has been discussed that these two theories 
can be understood as the same theory 
if the cutoff surface is placed slightly outside the horizon. 
The bulk viscosity is different for these two theories, but 
it does not contribute to physics since the fluid becomes incompressible 
in the near horizon limit. 
In the AdS/CFT correspondence, 
it is known that the fluid becomes compressible and the bulk viscosity is zero, 
apart from the near horizon limit. 
In this paper, we consider the Fluid/Gravity correspondence 
in asymptotically non-AdS geometries. 
We put the cutoff surface near 
but at a finite distance from the horizon. 
Then, the model becomes the membrane paradigm with compressible fluid. 
We show that the bulk viscosity is not negative at least within the linear response regime. 
We also discuss the higher derivative corrections in the stress-energy tensor. 
\end{abstract}

\baselineskip 18pt   

\end{titlepage}

\newpage

\section{Introduction}

According to the AdS/CFT correspondence 
\cite{Maldacena:1997re, Gubser:1998bc, Witten:1998qj, Witten:1998zw}, 
the anti-de Sitter (AdS) spacetime corresponds to the 
conformal field theory on the asymptotic boundary. 
At finite temperature and in the low frequency and long wavelength regime, 
a fluid appears in the conformal field theory 
and it can be seen in a black hole in the AdS spacetime 
\cite{Son:2002sd, Policastro:2002se, Policastro:2002tn, Bhattacharyya:2008jc}. 
This correspondence is known as the holographic hydrodynamics or the Fluid/Gravity correspondence. 
But before the discovery of the AdS/CFT correspondence, it has been discussed 
that a black hole can be described as a fictitious fluid surrounding the horizon. 
The earlier studies are known as the membrane paradigm 
\cite{Damour:1978cg, Damour:1982, Price:1986yy, Parikh:1997ma}. 

In the membrane paradigm, the stretched horizon, 
which is located slightly outside the horizon, is considered 
\cite{Price:1986yy, Parikh:1997ma}. 
From the viewpoint of an observer outside the black hole, 
the effects of the black hole are reproduced by matters on the stretched horizon. 
The fluid structure which corresponds to the black hole 
can be seen in the matters on the stretched horizon. 
A significant property of the fluid is the  negative bulk viscosity, 
which was expected to indicate the instability of the black hole. 

In the AdS/CFT correspondence, 
the fluid appears in the boundary of the AdS spacetime. 
According to the GKPW relations \cite{Gubser:1998bc, Witten:1998qj}, 
the boundary condition at the AdS boundary, $r\to\infty$, 
corresponds to the source term in the dual field theory side. 
In this sense, the fluid lives on the AdS boundary. 
In the AdS/CFT correspondence, the boundary of the AdS spacetime 
can be placed at a finite radius. 
The position of the boundary corresponds to 
the UV cutoff in the field theory side 
\cite{Susskind:1998dq, Akhmedov:1998vf, Alvarez:1998wr}. 
We call the boundary at the finite radius as the ``cutoff surface'' 
and take the Dirichlet boundary condition on the surface. 
The transport coefficients can be calculated 
by using the Kubo formula in the linear response theory 
\cite{Policastro:2002se, Policastro:2002tn}. 
Generally, the transport coefficients are obtained as functions of the position of the surface, 
and it can be interpreted as their scale dependence 
\cite{Iqbal:2008by, Bredberg:2010ky, Bredberg:2011jq}. 
The bulk viscosity in the case of the Schwarzschild-AdS black hole
is zero, which is independent of the position of the cutoff surface 
\cite{Kuperstein:2011fn, Brattan:2011my}. 
The Fluid/Gravity correspondences for black branes are 
also studied in \cite{Emparan:2012be, Emparan:2013ila}, 
in which the bulk viscosity is not zero but positive. 

Although the Fluid/Gravity correspondence and membrane paradigm 
are proposed in the different frameworks, 
it is expected that they are related to each other 
\cite{Bredberg:2010ky, Bredberg:2011jq, Brattan:2011my, Emparan:2013ila, Marolf:2012dr}. 
In fact, the stress-energy tensor is given by 
the Brown-York tensor \cite{Brown:1992br} in both theories. 
It has been shown that some transport coefficients agree when 
the cutoff surface approaches to the horizon \cite{Bredberg:2010ky}. 
Since the near horizon geometry of black holes is given by the Rindler space, 
it is sufficient to consider the Rindler space if 
the cutoff surface is at an infinitesimal distance from the horizon. 
The Fluid/Gravity correspondence in the Rindler space 
is studied in \cite{Bredberg:2011jq}. 
The relation between the sound modes in the Schwarzschild-AdS and Rindler space is 
discussed in detail in \cite{Brattan:2011my, Marolf:2012dr}. 
The fluid becomes incompressible 
for the Fluid/Gravity correspondence in the Rindler space. 
The bulk viscosity does not contribute to physics and 
can be arbitrary in the Rindler space, or equivalently, 
if the cutoff surface is very near the horizon. 
In fact, the bulk viscosity takes different values if 
the Rindler limit is taken in different geometries. 
In \cite{Matsuo:2012pi}, it has been proposed that 
it is not inconsistent because of this incompressibility 
that the bulk viscosity in the AdS/CFT correspondence 
does not agree with that in the membrane paradigm. 
In order to calculate the bulk viscosity, 
the cutoff surface must be sufficiently separated from the horizon, 
and we need to consider outside of the Rindler limit. 
Then, the fluid becomes compressible and the bulk viscosity 
contributes to physics of the fluid. 
In this sense, the zero bulk viscosity in the Fluid/Gravity correspondence 
is more physically meaningful than the negative bulk viscosity in the membrane paradigm. 
It should be noticed that the black holes do not have universal structure 
outside the Rindler limit, and hence we need to calculate 
the bulk viscosity for each black hole individually. 

The counter terms in the Fluid/Gravity correspondence corresponds to 
the effects of the geometry inside the stretched horizon in the membrane paradigm. 
They do not equal in general, but can be understood  
as the scheme dependence of the renormalization 
in the holographic renormalization group. 
Here, we consider only the cutoff surface near the horizon, 
and treat these two theories in the same framework. 

Since the negative bulk viscosity which was calculated in the membrane paradigm 
does not contribute to physics due to the incompressibility, 
it is interesting to consider what is the bulk viscosity which contribute to the physics. 
In the case of the Schwarzschild-AdS spacetime, the bulk viscosity was calculated 
on the cutoff surface at arbitrary radius, and is know to be zero. 
It was also calculated in the black brane solutions \cite{Emparan:2012be, Emparan:2013ila}.
However, the fluid structure does not appear at arbitrary radius 
for the most of the black holes. 
Even in this case, as has been discussed in the membrane paradigm, 
the black holes correspond to fluids as long as the cutoff surface is near the horizon. 
In particular, the membrane paradigm can be applied to the black holes 
which is not asymptotically AdS nor based on string theory. 

In this paper, we generalize the previous studies and 
consider the sound modes in black holes 
which are not asymptotically AdS and have compact horizons. 
Since the fluid becomes incompressible in the Rindler limit, 
we put the cutoff surface near but at a finite distance from the horizon. 
Since the black holes do not have universal characteristics here, 
we focus on the simplest cases, 
with the (asymptotically flat) Schwarzschild black hole in mind. 
We show that the fluid structure can be seen even outside the Rindler limit. 
We calculate the transport coefficients within the linear response regime, 
and show that the bulk viscosity is not negative. 

This paper is organized as follows.
In section~\ref{sec:fluid}, we consider a fluid on a maximally symmetric space and describe the expression of the linear response of the stress-energy tensor for the sound modes.
In section~\ref{sec:black}, we describe the linear response of the Brown-York tensor in a maximally symmetric  black hole and compare it with that of the stress-energy tensor of the fluid on the maximally symmetric space.
In section~\ref{sec:fluidgravity}, we briefly review  the Fluid/Gravity correspondence 
and membrane paradigm. 
In section \ref{sec:pertblack},  we consider the metric perturbations 
of the maximally symmetric black holes. 
In  section \ref{sec:emcutoff}, we calculate the Brown-York tensor 
on the cut-off surface at $r=r_c$, 
and compare it with the fluid stress-energy tensor. 
In section~\ref{sec:HigherDeriv}, we consider 
possible corrections in the fluid stress-energy tensor 
which reproduce the higher derivative terms in the Brown-York tensor 
in the gravity side. 
In section~\ref{sec:CounterTerm}, 
we make a few comments on the counter terms and 
contribution from the inside geometry to the junction condition. 
Section~\ref{sec:conclusion} is devoted to the conclusion and discussions. 
In the Appendix, we write down the detailed expressions which are used in our calculation.

\section{Linear response of fluid on maximally symmetric space} \label{sec:fluid}
We consider a fluid on a $n$-dimensional maximally symmetric space $\mathcal K_n$. 
We take the background metric as
\begin{align}
ds^2=-dt^2+d\sigma_n^2, \label{bgmetric}
\end{align}
where 
\begin{align}
 d \sigma_n^2 = \hat\gamma_{ij}(z)dz^idz^j , ~~~~~~~(i,j=1,\cdots, n), \label{smetric}
\end{align}
is the metric on $\mathcal K_n$. 
We denote the sectional curvature and covariant derivative 
on $\mathcal K_n$ as $K$ and $\hat D_i$, respectively. 

To obtain the expressions of the linear response of the stress-energy tensor, we apply the metric perturbations to the fluid. 
We denote the metric perturbations as 
\begin{align}
g_{\mu\nu}&=\gbar_{\mu\nu}+\delta g_{\mu\nu}, 
\end{align}
where $\bar g_{\mu\nu}$ is the background metric \eqref{bgmetric}. 
In general, tensors with at most rank two on the maximally symmetric space 
can be decomposed into scalar, vector, and tensor types and 
can be expanded in terms of the harmonic functions on $\mathcal{K}_n$. 
In this paper, we focus on the scalar type. 
The metric perturbations can be expanded as
\begin{subequations}\label{metpertfl}
\begin{align}
\delta g_{00}&=-f_{0}^0\,e^{-i \omega t}\, \Sbb,  \\
\delta g_{0i}&=f_0\,e^{-i \omega t}\, \Sbb_i,  \\
\delta g_{ij}&=2 e^{-i \omega t}(H_L\hat\gamma_{ij}\Sbb+H_T\Sbb_{ij}),
\end{align}
\end{subequations}
where the coefficients $f^0_0, f_0,H_L$, and $H_T$ are functions of $\omega$ and $k$, 
and $\Sbb$, $\Sbb_i$, and $\Sbb_{ij}$ are the harmonic functions 
of the scalar type 
which are characterized by the momentum $k$. 
Hereafter, we will omit $k$ and $\omega$ dependences in 
the harmonic functions and the coefficients. 
The harmonic functions obey the following formulas: 
\begin{align}
\Dhat^i\Dhat_i \Sbb=-k^2 \Sbb, 
\end{align}
for the scalar harmonics $\Sbb$, 
\begin{align}
\Sbb_i&=-\frac{1}{k}\Dhat_i \Sbb, \\
\Dhat_i\Sbb^i&=k\Sbb, \\
\Dhat_i\Sbb_j&=\Dhat_j\Sbb_i, \\
\Dhat^i\Dhat_i \Sbb_j&=((n-1)K-k^2)\Sbb_j,
\end{align}
for the vector harmonics $\Sbb_i$, and 
\begin{align}
\Sbb_{ij}&=\frac{1}{k^2}\Dhat_i\Dhat_j\Sbb+\frac{1}{n}\hat\gamma_{ij}\Sbb, \notag \\
&=-\frac{1}{k}\Dhat_i\Sbb_j+\frac{1}{n}\hat\gamma_{ij}\Sbb, \\
\Sbb^i_i&=0, ~~~~~~\Sbb_{ij}=\Sbb_{ji}, \\
\Dhat_i\Sbb^i_j&=\frac{n-1}{n}\frac{k^2-nK}{k}\Sbb_j, \\
\Dhat^i\Dhat_i\Sbb_{jk}&=(2nK-k^2)\Sbb_{jk},
\end{align}
for the tensor harmonics $\Sbb_{ij}$. 
For $K=1$, $k^2$ takes discrete values,
\begin{align}
k^2=l(l+n-1),~~~~~~~l=0,1,2,\cdots,
\end{align}
while for $K\leq 0$, $k^2$ takes any non-negative real value.

In general, the stress-energy tensor for a fluid is expressed as 
\begin{equation}
 T^{\mu\nu} = (\varepsilon + P) u^\mu u^\nu + P g^{\mu\nu} + \tau^{\mu\nu} , \label{FluidStress}
\end{equation}
where $\varepsilon$, $P$, and, $u^\mu$ are 
the energy density, pressure, and velocity field, respectively. 
The velocity field satisfies the normalization condition $u^{\mu}u_{\mu} = -1$. 
In the first-order formalism of  fluid mechanics,
the viscous stress tensor $\tau^{\mu\nu}$ is given by 
\begin{align}
 \tau^{\mu\nu} 
 &= 
 - 2 \eta \sigma^{\mu\nu} 
 - \zeta \theta \Delta^{\mu\nu} , \label{ViscousStress}
\end{align}
where $\eta$ and $\zeta$ are the shear viscosity and bulk viscosity, respectively, and
\begin{align}
 \Delta^{\mu\nu} 
 &= 
 u^\mu u^\nu + g^{\mu\nu} , \label{Projection}
\\
 \sigma^{\mu\nu} 
 &= \frac{1}{2}
 \Delta^{\mu\alpha} \Delta^{\nu\beta} 
 \left(
  \nabla_\alpha u_\beta + \nabla_\beta u_\alpha 
  - \frac{2}{n} g_{\alpha\beta} \nabla_\gamma u^\gamma
 \right) , \label{Shear}
\\
 \theta &= \nabla_\alpha u^\alpha . \label{Expansion}
\end{align}

By introducing the metric perturbations \eqref{metpertfl} in the fluid, 
the energy density, pressure, and velocity field receive 
the linear responses as 
\begin{align}
 \varepsilon 
 &= 
 \bar\varepsilon + \delta \varepsilon\,e^{-i \omega t}\,\Sbb , 
\\
 P 
 &= 
 \bar P + \delta P\,e^{-i \omega t}\,\Sbb , 
\\
 u^\mu 
 &= 
 \bar u^\mu + \delta u^\mu ,
\end{align}
where $\delta \varepsilon$ and $\delta P$ are functions of $\omega$ and $k$.
Now, we consider the response of the fluid at rest and hence 
we take 
\begin{equation}
 \bar u^\mu = \delta^\mu_0 .
\end{equation}
Then,  the response of the velocity field takes the form of 
\begin{align}
\delta u^0 &=-\frac{1}{2}f^0_0\,e^{-i \omega t}\, \Sbb, \\
\delta u^i &=u\,e^{-i \omega t} \,\Sbb^i,
\end{align}
at the linear order of the metric perturbations. Here, $u$ is a function  of $\omega$ and $k$.
Thus, the responses of the stress-energy tensor are expressed as 
\begin{align}
\delta T^0{}_0&=-\delta \varepsilon\, e^{-i \omega t} \Sbb, \label{eq:hydroT00} \\
\delta T^i{}_0&=-(\ebar +\pbar)u\, e^{-i \omega t} \Sbb^i, \label{eq:hydroT0i}\\
\delta T_L 
 &=[\delta P -\zeta (ku-i\omega nH_L)] \ , \\
\delta T_T 
 &=2\eta (ku+i\omega H_T) \ ,
\label{eq:hydroTij}
\end{align}
where $\delta T_L$ and $\delta T_T$ are the trace part and traceless part of 
the spatial components $\delta {T^i}_j$: 
\begin{equation}
 \delta T^i{}_j = \delta T_L\, e^{-i \omega t} \delta^i{}_j + \delta T_T\, e^{-i \omega t}  \Sbb^i{}_j.
\end{equation}
From the conservation law $\nabla_\mu T^\mu{}_\nu = 0$, we obtain
\begin{align}
 0 
 &= 
 i \omega \delta \varepsilon 
 - (\bar \varepsilon + \bar P ) (ku - i n \omega H_L), \label{eq:cons1}
\\
 0 
 &= 
 - k c_s^2\delta \varepsilon - i \omega (\bar \varepsilon + \bar P ) u  
 + 2 \eta \frac{n-1}{n}(k^2- n K) u + \zeta k^2 u 
\notag\\&\qquad
 - \frac{1}{2} k (\bar \varepsilon + \bar P ) f^0_0 
 -i \omega (\bar \varepsilon + \bar P )f_0 
 -i \zeta n \omega k H_L + 2 i \eta \frac{n-1}{n} \frac{k^2-nK}{k}\omega H_T , \label{eq:cons2}
\end{align}
where we have used
\begin{equation}
 \delta P = c_s^2 \delta \varepsilon ,
\end{equation}
and $c_s^2$ is the velocity of sound.
Solving (\ref{eq:cons1}) and (\ref{eq:cons2}) for $u$ and $\delta \varepsilon$, we find
\begin{align}
u&=\frac{i\omega}{k} \frac{-\frac{1}{2}k^2f^0_0-i\omega k f_0+\frac{2(n-1)}{n}(k^2-nK)i\omega \etahat H_T+nk^2(c_s^2-i\omega \zetahat)H_L}{-\omega^2+c_s^2 k^2-i\omega (\Gamma_s k^2 -2(n-1) K\etahat)}, 
\label{solu}\\
\delta \varepsilon &=(\ebar+\pbar)\bigg(\frac{k}{i\omega}u-nH_L\bigg) \notag \\
&=(\ebar+\pbar)\frac{-\frac{1}{2}k^2f^0_0-i\omega k f_0+\frac{2(n-1)}{n}(k^2-nK)i\omega \etahat H_T+n\big(\frac{2(n-1)}{n}(k^2-nK)i\omega \etahat +\omega^2\big)H_L}{-\omega^2+c_s^2 k^2-i\omega (\Gamma_s k^2 -2(n-1) K\etahat)}, \label{sole}
\end{align}
where
\begin{align}
\etahat&=\frac{\eta}{\ebar+\pbar}, \\
\zetahat&=\frac{\zeta}{\ebar+\pbar}, \\
\Gamma_s&=\frac{2(n-1)}{n}\etahat+\zetahat . 
\end{align}
The sound pole 
\begin{equation}
 \Delta_s = -\omega^2+c_s^2 k^2-i\omega (\Gamma_s k^2 -2(n-1) K\etahat) = 0 \label{FluidSoundPole}
\end{equation}
can be read off from the denominator of (\ref{solu}) or (\ref{sole}). 

Substituting the solutions of the conservation law \eqref{solu} and \eqref{sole} into (\ref{eq:hydroT00})-(\ref{eq:hydroTij}), 
 the stress-energy tensor takes the form of  
\begin{align}
 \delta T_I 
 &= 
 \frac{\mathcal N_I}{\Delta_s} , 
\end{align}
where the numerators $\mathcal{N}_I=(\mathcal N^0{}_0, \mathcal N_0, \mathcal N_L, \mathcal N_T)$ of the components  of the stress-energy tensor $\delta T_I=(T^0{}_0, T_0, T_L, T_T)$
are obtained as 
\begin{subequations}\label{FluidNum}
\begin{align}
 \mathcal N^0{}_0 
 &=
 \frac{k^2}{2}(\bar \varepsilon + \bar P )  f^0_0
 +i k \omega (\bar \varepsilon + \bar P ) f_0
\notag\\&\qquad
 + \omega  (\bar \varepsilon + \bar P ) \left[-n \omega +2 i \hat\eta  (n-1) (K n-k^2 )\right] H_L
 - 2 i \eta \omega \frac{n-1}{n}(k^2-nK) H_T \ ,
\\
 \mathcal N_0 
 &=
  \frac{1}{2} i k \omega (\bar \varepsilon + \bar P ) f^0_0
 -\omega ^2 (\bar \varepsilon + \bar P ) f_0
\notag\\&\qquad
 -i n \omega k (\bar \varepsilon + \bar P ) (c_s^2-i \omega\hat \zeta) H_L
 + 2 \eta \omega ^2\frac{n-1}{n} \frac{k^2- nK}{k} H_T \ ,
\\
 \mathcal N_L 
 &=
 -\frac{1}{2} k^2 (\bar \varepsilon + \bar P ) ( c_s^2-i \omega \hat \zeta) f^0_0
 -i k \omega (\bar \varepsilon + \bar P ) (c_s^2-i \omega \hat\zeta) f_0
\notag\\&\qquad
 +  \omega (\bar \varepsilon + \bar P ) (c_s^2-i \hat \zeta  \omega ) 
 \left[n \omega +2 i \hat \eta  (n-1) (k^2-K n)\right] H_L
\notag\\&\qquad
 + 2 \eta \omega  \frac{n-1}{n}(k^2-nK) (\hat \zeta  \omega +i c_s^2) H_T \ ,
\\
 \mathcal N_T 
 &=
 -i \eta  k^2 \omega f^0_0
 + 2 \eta  k \omega ^2 f_0
 + 2 \eta  k^2 n \omega (\hat \zeta  \omega +i c_s^2) H_L
 + 2 i \eta  \omega (-\omega^2 + c_s^2 k^2 -i \hat\zeta \omega k^2) H_T \ .
\end{align}
\end{subequations}

\section{Linear response of maximally symmetric black holes} \label{sec:black}

In this section, we consider the linear response theory in the gravity side. 
We consider the black hole geometries whose boundary is the geometry which 
we have discussed in the previous section. 
Then, we introduce the scalar type perturbations to the metric 
and calculate the linear response in the energy-momentum tensor on the boundary. 
The transport coefficients are calculated at the leading order of the near horizon expansion. 
Since the corresponding fluid becomes incompressible in the Rindler space, 
the leading order contributions for the sound modes give the corrections to the Rindler limit.

\subsection{Fluid/Gravity correspondence and membrane paradigm} \label{sec:fluidgravity}

The correspondence between the black holes and fluids 
have been discussed in two different frameworks, 
the Fluid/Gravity correspondence and membrane paradigm. 
These two theories are related to each other 
by using the holographic renormalization group. 
The stress-energy tensor is given by the same formula 
in both theories up to terms which can be interpreted as 
the scheme dependence of renormalization. 
We first explain the Fluid/Gravity correspondence, in particular 
focusing on the stress-energy tensor of the fluid. 
Then, we discuss the relation between 
the Fluid/Gravity correspondence and membrane paradigm. 

In the AdS/CFT correspondence, 
the energy-momentum tensor on the boundary is given by the Brown-York tensor, 
which is introduced to define a quasi-local energy. 
By using the Brown-York tensor, the quasi-local charge $Q_\xi$ associated 
to a Killing vector $\xi$ is expressed as
\begin{equation}
 Q_\xi = \int d^d x \sqrt{-\gamma}\, T^{0\mu}\xi_\mu \ . 
\end{equation}
The Brown-York tensor is defined by 
\begin{equation}
 T^{\mu\nu} 
 = \frac{2}{\sqrt{-\gamma}}\frac{\delta S_\text{grav}}{\delta \gamma_{\mu\nu}} \ ,  \label{BYdef}
\end{equation}
where $S_\text{grav}$ is the gravitational action and $\gamma_{\mu\nu}$ is the induced metric on the boundary. 
The action is evaluated at the classical configuration of the metric. The induced metric on the boundary plays 
the role of the boundary condition of the metric. 
The definition of the Brown-York tensor \eqref{BYdef} is 
consistent to the GKPW relation for the energy-momentum tensor \cite{Balasubramanian:1999re}. 

We consider the $(d+1)$-dimensional Einstein-Hilbert action for the gravity theory: 
\begin{equation}
 S_\text{grav} 
 = 
 \frac{1}{16\pi G} \int d^{d+1} x \sqrt{-g} (R - 2 \Lambda) 
 + \frac{1}{8\pi G} \int_{r=r_c} d^d x \sqrt{-\gamma} \mathcal K \ ,
\end{equation}
where $R$ and $\Lambda$ are the scalar curvature and cosmological constant 
in the bulk and $\mathcal K$ is the trace of 
the extrinsic curvature $\mathcal K_{\mu\nu}$ on the cutoff surface $r=r_c$. 
Although the ordinary AdS boundary is located at $r\to\infty$, 
we consider only inside of the cutoff surface and impose the boundary conditions there. 
Then, the Brown-York tensor is expressed in terms of the extrinsic curvature as 
\begin{equation}
 T^{\mu\nu} 
 = 
 \frac{1}{8\pi G} \left(\mathcal K^{\mu\nu} - \gamma^{\mu\nu} \mathcal K\right) \ .  \label{eq:adsby}
\end{equation}
In order to obtain well-defined charges, 
we should further define the reference geometry and subtract its effect from 
the above definition, or add appropriate counter terms instead. 
In the AdS/CFT correspondence, the counter terms are determined 
such that they are written in terms of covariant quantities, 
and become proportional to the induced metric or 
Einstein tensor on the boundary \cite{Balasubramanian:1999re}. 

In the membrane paradigm, a black hole is replaced by 
a fictitious fluid on the stretched horizon, 
which is a timelike hypersurface slightly outside the event horizon 
\cite{Price:1986yy, Parikh:1997ma}. 
Classically, an observer who remains outside the black hole
are not affected by the dynamics inside the black hole. 
Then, the domain of  integration of the action can be restricted to the external of the black hole, 
but the appropriate surface terms on the horizon must be added into the action. 
These surface terms can be interpreted as the effects of the matter on the stretched horizon. 
Since this matter behaves as a fluid, the black hole can be replaced by the fluid on the stretched horizon. 
 
The stress-energy tensor of the fluid is derived by using 
the Israel junction condition \cite{Israel:1966rt} 
and expressed in terms of the extrinsic curvature as 
\begin{equation}
 T^{\mu\nu} 
 = 
 \frac{1}{8\pi G} \left(\mathcal K^{\mu\nu}_+ - \gamma^{\mu\nu} \mathcal K_+\right) 
 - \frac{1}{8\pi G} \left(\mathcal K^{\mu\nu}_- - \gamma^{\mu\nu} \mathcal K_-\right) \ ,  
\end{equation}
where $\mathcal K^{\mu\nu}_+$ and $\mathcal K^{\mu\nu}_-$ are the extrinsic curvatures 
of the geometry outside and inside the stretched horizon, respectively. 
In order to describe all effects of black hole by the matters on the stretched horizon, 
the  geometry inside the black hole is usually taken to be  flat spacetime. Then, the stress-energy tensor in the membrane paradigm takes the same form as (\ref{eq:adsby}).

Although both the Fluid/Gravity correspondence and membrane paradigm 
 describe the physics of black holes in terms of fluids, there are some differences between them. Firstly,  in the Fluid/Gravity correspondence, 
the fluid lives on the boundary of AdS spacetime
while  in the membrane paradigm, the fluid is placed on the stretched horizon. 
This discrepancy can be resolved by using the holographic renormalization group. 
In the AdS/CFT correspondence, the boundary of the AdS spacetime is 
originally placed at $r=r_c\to\infty$. 
A  finite but large $r_c$ can be interpreted as 
an UV cut-off in the CFT side. 
Lowering $r_c$ corresponds to the renormalization group flow in CFT. 
Therefore, the membrane paradigm can be understood 
as the IR limit of the Fluid/Gravity correspondence.

Secondly, the AdS/CFT correspondence deals with the black hole geometry inside the cut-off surface, 
while the  geometry inside the stretched horizon is replaced 
by the flat spacetime in the membrane paradigm. 
However, the membrane paradigm can also be interpreted as 
the correspondence between two different geometries; 
one is the black hole geometry all over the spacetime, 
and the other is given by connecting black hole geometry and flat space at the stretched horizon. 
Since the outside is given by the black hole geometry in both sides, 
the membrane paradigm is the correspondence 
between two models inside the stretched horizon; 
one is the black hole geometry and the other is the flat spacetime with fictitious fluid. 
In this sense, the membrane paradigm also deals with the black hole geometry inside the cutoff surface. 
Therefore, the Fluid/Gravity correspondence and membrane paradigm 
is equivalent when  the cut-off surface is near the horizon. 

It should be also noticed that the identification 
of the fluid variables is different for the Fluid/Gravity correspondence and membrane paradigm. 
In the membrane paradigm, the normal vector of the horizon becomes 
tangent to the horizon and is the only vector which is pointing in the causal direction on the horizon. 
Therefore, the normal vector is identified to the velocity field of the fluid. 
However, this is valid only in the limit in which the cutoff surface agrees with the true horizon, 
and cannot be applied to the AdS/CFT correspondence 
in which the cutoff surface can be placed at an arbitrary radius. 
However, the Einstein equation contains the constraints which corresponds 
to the continuity equation and Navier-Stokes equation. 
Then, the fluid variables can be solved by using the constraints 
and they are expressed in terms of the induced metric on the surface, 
as we have seen in the previous section. 
After imposing the constraints, the stress-energy tensor of the fluid   in the Fluid/Gravity correspondence
becomes the same  as that of the  membrane paradigm as long as 
the corresponding solution of gravity theory is the same. 
In order to solve the Einstein equation, 
we have to impose the ingoing boundary condition at the horizon for the propagating modes. 
The other boundary conditions are arbitrary. 
As we will see later, the Dirichlet conditions on the cutoff surface 
give the most general solutions for the linear perturbations. 
Therefore, fluids which correspond to solutions of the Einstein equation 
can always be parametrized by the induced metric on the cutoff surface, 
at least in the linear response regime.

\subsection{Metric perturbations of black hole} \label{sec:pertblack}

The metric of the maximally symmetric black hole in $(n+2)$-dimensional spacetime is given by 
\cite{Kodama:2003jz}, 
\begin{equation}
 ds^2 = -f(r)dt^2 + \frac{dr^2}{f(r)} + r^2 d \sigma_n^2 \ , \label{gmetric}
\end{equation}
where 
\begin{equation}
 f(r) = K - \frac{2M}{r^{n-1}} - \lambda r^2 \ . 
\end{equation}
Here, $M$ is related to the black hole mass, 
and $\lambda$ is related to the cosmological constant $\Lambda$ by 
\begin{equation}
 \lambda = \frac{2\Lambda}{n(n+1)} \ .
\end{equation}
The spatial metric $d \sigma_n^2$ is that of $\mathcal K_n$, 
which is given by \eqref{smetric}, 
and $K$ is the sectional curvature on it. 

At the horizon $r=r_H$, $f(r)$ satisfies $f(r_H)=0$.
The Hawking temperature is given by 
\begin{equation}
 T_H = \frac{1}{4\pi} f'(r_H) . 
\end{equation}

We introduce scalar type perturbations to the metric \eqref{gmetric}. 
As in the previous section, the metric perturbations can be expanded in terms of 
the harmonic functions on $\mathcal K_n$. 
We choose the gauge such that the perturbations with $r$-component vanish: 
\begin{equation}
 \delta g_{\mu r} = 0 . 
\end{equation}
Then, the metric perturbations can be expressed as \cite{Kodama:2003jz}, 
\begin{subequations}\label{metpertgr}
\begin{align}
\delta g_{00}&= - f(r) f_{0}^0(r) \,e^{-i \omega t}\, \Sbb,  \\
\delta g_{0i}&= r f_0(r) \,e^{-i \omega t}\, \Sbb_i,  \\
\delta g_{ij}&=2 r^2 e^{-i \omega t}\left(H_L(r) \hat\gamma_{ij}\Sbb+H_T(r) \Sbb_{ij}\right).
\end{align}
\end{subequations}
Since the Einstein tensor $G_{MN}$ has the same structure  as the metric perturbations
under the expansion in terms of the harmonic functions, 
the relevant equations are  the following seven components of the Einstein tensor:
\begin{align}
 E_{tt} &= 0 \ , &
 E_{tr} &= 0 \ , &
 E_{rr} &= 0 \ , \\
 E_{t}  &= 0 \ , &
 E_{r}  &= 0 \ , &
 E_L    &= 0 \ , &
 E_T    &= 0 \ ,
\end{align}
where 
\begin{equation}
 E_{\mu\nu} = G_{\mu\nu} + \Lambda g_{\mu\nu},
\end{equation}
and it can be expanded in terms of the spherical harmonics as 
\begin{align}
 E_{ti} &= E_{t} \Sbb_i \ , &
 E_{ri} &= E_{r} \Sbb_i \ , &
 E_{ij} &= E_L \hat\gamma_{ij} \Sbb + E_T \Sbb_{ij} \ .  
\end{align}
These seven equations can be separated into 
four second-order differential equations with respect to $r$, 
and three first-order differential equations. 
Since the Einstein tensor satisfies the Bianchi identity, 
these equations are related to each other. 
The independent equations are 
one second-order differential equation and 
three first-order differential equations. 
The first-order differential equations, which are constraint equations, 
give constraints on the boundary conditions of the metric perturbations 
and the second-order differential equation describes the propagation of the sound mode. 

These differential equations provide in total five integration constants in the solution. 
One of them is fixed by 
imposing the incoming wave boundary condition at the horizon for the propagating mode.
For the  remaining  four integration constants, 
we impose the Dirichlet boundary conditions at the cutoff surface $r=r_c$ 
on the metric perturbations $f^0_0$, $f_0$, $H_L$, and $H_T$. 
Then, all integration constants are determined. 

By taking a specific combination of the metric perturbations, 
we obtain a second-order differential equation only in terms of the combination 
\cite{Kodama:2000fa, Kodama:2003jz}. 
This combination and the equation are referred to 
as the master field and master equation, respectively. 
The combination for the master equation is not unique. 
Here, we take the master field as
\begin{align}
 \Phi(r) &= 2 r^{n/2} \left\{H_T(r) 
 -\frac{i n \left[f(r) f_0'(r) - r (f(r)/r)' f_0(r)+2 i \omega  H_L(r)\right]}
   {\omega  \left( 2k^2+ n r f'(r)-2 n f(r)\right)} \right\} \ . \label{MasterField}
\end{align}
Then, the master equation is 
\begin{equation}
 \frac{d}{dr}\left(f(r)\frac{d\Phi(r)}{dr}\right) 
 + \frac{\omega^2}{f(r)} \Phi(r) - V(r) \Phi(r)  = 0 \ , \label{MasterEq}
\end{equation}
where the pontential $V(r)$ is 
\begin{equation}
 V(r) = \frac{Q(r)}{16 r^2 H^2(r)} \ ,
\end{equation}
and
\begin{align}
 H(r) 
 &= 
 k^2 - n f(r) + \frac{1}{2} n r f'(r) \ ,
\\
 Q(r) 
 &= 
 4 M^2 n (n+1) r^{2-2 n} \left[4 k^2 (2 n^2-3 n+4)+K n (n^3-13 n^2+14
 n-8)\right] 
\notag\\&\qquad
 +\lambda  r^2 
   \Bigl[24 M (n-2) n^2 (n+1) r^{1-n} 
    (k^2-K n) -4 (n-4) (n-2) (k^2-K n)^2
\notag\\&\qquad\qquad\qquad
 -4 M^2 n^3 (n+1)^2 (n+2) r^{2-2n}
   \Bigr]
\notag\\&\qquad
 -24 M n r^{1-n} (k^2-K n) \left[k^2 (n-4)+K n (n^2-2n+2)\right]
\notag\\&\qquad
 +16 (k^2-K n)^3+4 K n (n+2) (k^2-K n)^2+8M^3 n^4 (n+1)^2 r^{3-3 n} \ . 
\end{align}

The solution of $\Phi$ generally takes the following form near the horizon: 
\begin{equation}
 \Phi 
 \sim 
 C_0 (r-r_H)^{-i\omega/f'(r_H)} 
 + \bar C_0 (r-r_H)^{i\omega/f'(r_H)} \ , \label{eq:phi}
\end{equation}
where $C_0$ and $\bar C_0$ are the integration constants.
The first term of (\ref{eq:phi}) corresponds to the ingoing mode and 
the second term corresponds to the outgoing mode. 
We take the ingoing boundary condition $\bar C_0 =0$. 

In the case of the AdS spacetime, 
the structure of fluid appears in the hydrodynamic regime, $\omega, k\to 0$,
on the cutoff surface with an arbitrary $r_c$. 
The master equation can be solved by expanding the master field for small $\omega$ and $k$. 
However, $k$ takes only discrete values for $K>0$, and we cannot take $k\to 0$ limit. 
Generally, the fluid structure appears when the wavelength is much longer than the inverse of  temperature. 
Since we consider the matters on the cutoff surface, we should take the blueshift into account. 
By using the local quantities, the condition is given by 
\begin{align}
 \frac{k}{r_H} &\ll \frac{T_H}{\sqrt{f}} \ , &
 \frac{\omega}{\sqrt{f}} &\ll \frac{T_H}{\sqrt{f}} \ .  
\end{align}
For $K>0$, the wavelength is bounded above by the horizon radius, and 
hence, $k$ cannot be arbitrarily small. 
However, the fluid structure appears near the horizon 
since the blueshift factor, $1/\sqrt{f}$, becomes very large.  
Since the wavelength can be same order to the horizon radius, 
the condition can be expressed in terms of lengths in the black hole geometries as  
\begin{equation}
 r_c - r_H \ll r_H \ , 
\end{equation}
for $\Lambda=0$. For a non-zero cosmological constant, 
the condition becomes stronger but can be satisfied by 
putting the cutoff surface sufficiently near the horizon. 
Thus, for example, in the case of asymptotically flat spacetime, 
the fluid structure appears only near the horizon. 
Here, we focus on the cut-off surface near the horizon, 
and calculate the Brown-York tensor in expansion around the horizon $r=r_H$. 

Near the horizon, the master field $\Phi$ can be expanded in terms of $(r-r_H)$ as 
\begin{equation}
 \Phi(r) = C_0 f(r)^{-i\omega/f'(r_H)}\left(1+C_1 f(r) + \mathcal O((r-r_H)^2)\right). \label{eq:masterp}
\end{equation}
By solving the master equation, $C_1$ is obtained as 
\begin{align}
 C_1&= 
 \frac{1}{2 r^2 f^{\prime\,4}(2k^2 + n r f')} 
\notag\\&\quad
 \times\Bigl\{
+f^{\prime\,2}
   \left[-4 r f' \left(k^2 (n-2)-K (n-1) n\right)+n^2 r^2 f^{\prime\,2}+4 k^2
   \left(k^2-2 K (n-1)\right)\right]
\notag\\&\quad\qquad
+4 i \omega  f' \left[ r f' \left(K (n-1) n-3 k^2
   (n-2)\right)+n r^2 f^{\prime\,2}+2 k^2 \left[k^2-3 K (n-1)\right)\right]
\notag\\&\quad\qquad
 +4 \omega ^2 \left[4 r f' \left(k^2 (n-2)-K (n-1) n\right)-n^2 r^2
   f^{\prime\,2}-4 k^2 \left(k^2-2 K (n-1)\right)\right]
\Bigr\}\biggr|_{r=r_H} \ .
\end{align}
It is straightforward to calculate 
the higher order corrections of the near horizon expansion. 
It should be noticed that 
we have not imposed the condition of $k \ll 1$ or $\omega \ll 1$. 

\subsection{Fluid stress-energy tensor from gravity} \label{sec:emcutoff}

For the background geometry \eqref{gmetric}, 
the Brown-York tensor is obtained as 
\begin{subequations}\label{bgBY}
\begin{align}
\bar{T}^0{}_0&=nf^{1/2}r^{-1}, \label{eq:zerot00}\\
\bar{T}^i{}_0&=0, \\
\bar{T}_L &=\frac{1}{2}f^{-1/2}(f'+2(n-1)fr^{-1}) .\label{eq:zerotij}
\end{align}
\end{subequations}
where 
\begin{equation}
\bar{T}^i{}_j = \bar T_L \delta^i_j. 
\end{equation}
Hereafter, we take $8\pi G = 1$. 
By introducing the perturbations of the metric \eqref{metpertgr}, 
the linear responses of the Brown-York tensor are
\begin{align}
\delta T^0{}_0&=nf^{1/2}H_L' e^{-i\omega t} \Sbb,  \label{eq:BYT00} \\
\delta T^i{}_0&=-\frac{1}{2}f^{1/2}r^{-1}(r^{-1}f_t-f^{-1}f'f_t+f_t') e^{-i\omega t} \Sbb^i, \label{eq:BYT0i}\\
\delta T_L &= \frac{1}{2}f^{1/2}(f^{t'}_t+2(n-1)H_L') \ ,\\
\delta T_T 
&=-f^{1/2}H_T'\ , 
\end{align}
where
\begin{equation}
 \delta T^i{}_j = \delta T_L e^{-i\omega t}\delta^i{}_j + \delta T_T e^{-i\omega t}\Sbb^i{}_j . 
\end{equation}

The solution for the metric perturbations can be obtained 
by using the solution of the master equation \eqref{eq:masterp} 
and solving constraint equation. 
However, in order to calculate the Brown-York tensor, 
the full solution of the metric perturbations is not necessary 
but it is only needed to rewrite the first derivatives of the metric perturbations 
in terms of their boundary conditions on the cutoff surface. 
By using three constraint equations and the definition of the master field \eqref{MasterField}, 
we obtain the following relation: 
\begin{equation}
 h_I' = \mathcal A_{IJ} h_J + \mathcal B_I \Phi ,\label{FirstDeriv}
\end{equation}
where $h_I$ stands for the metric perturbations $f^0_0$, $f_0$, $H_L$, and $H_T$. 
Since these relations are complicated, we write the detailed expressions   in the \ref{apdx:fullex}. 
We rewrite the master field (\ref{eq:masterp}) as
\begin{equation}
 \Phi(r) = C F(r) \ , \label{eq:cf}
\end{equation}
where $F(r)$ is normalized as 
\begin{equation}
 F(r_c) = 1 \ . 
\end{equation}
The  constant $C$ can be fixed 
by imposing the Dirichlet boundary conditions on the metric perturbations \eqref{metpertgr}  at $r=r_c$. 

Since the definition of the master field \eqref{MasterField} contains 
a first derivative of the metric perturbations, 
the master equation is the third order differential equation 
when it is expressed in terms of the original variables. 
There is an additional integration constant 
in the constraint equations and the master equations. 
It can be fixed by the original second order differential equation, 
which can be expressed in terms of $h_I$, $h_I'$, $\Phi$ and $\Phi'$. 
Then, by using \eqref{eq:masterp} and \eqref{FirstDeriv}, 
we obtain a relation between the integration constant $C$ and Dirichlet boundary conditions as 
\begin{equation}
 C= \left.\frac{C_N}{F'(r)-C_D}\right|_{r=r_c} \ ,
\end{equation}
where 
\begin{align}
 C_D &= -\frac{k^2}{n r f(r)}
\notag
\\&\qquad
 + \frac{(2k^2 +n r f'(r)-n (n+1) f(r)) \left[k^2 ((n-1) f(r)- r f'(r))+2
   n r^2 \omega ^2\right]}{4 (n-1) n r f(r)^2 \left(k^2-n K\right)}
\notag
\\&\qquad
 + \frac{2 k^2 (n-2)n - (4 n - 4)K +n^2 [(n+1) f(r)-r f'(r)]}{2 r \left(2 k^2+n r
   f'(r)-n (n+1) f(r)\right)} \ ,
\displaybreak[1]
\\
 C_N &= 
-\frac{r^{{n}/{2}-1} (2 k^2 + n r f'(r) - n (n+1) f(r))}
 {2 (n-1) n f(r)^2 k^2 \left(k^2 - nK\right)}
\notag
\\&\qquad\times
 \Bigl\{n k^2 f(r) f^0_0 
 +2 i n r k \omega f_0 
 + n k^2 \left((n-1) f(r)-r f'(r)\right)H_L 
\notag
\\&\qquad\qquad
 + \left(-k^2 r f'(r)+k^2 (n-1) f(r)+2 n r^2 \omega ^2\right) H_T 
 \Bigr\} \ ,
\displaybreak[1]
\\
 F(r) &= 
 \frac{1+C_1 f(r)}{1+C_1 f(r_c)}\left(\frac{f(r)}{f(r_c)}\right)^{-i\omega/f'(r_c)} \ .
\end{align}
Substituting $\Phi(r_c) = C$ with these expressions into \eqref{FirstDeriv}, 
we obtain the expressions of the first derivatives of the metric perturbations 
in terms of their boundary conditions. 

The Brown-York tensor is given in the form of 
\begin{equation}
 \delta T_I = \left.\frac{\mathcal N_I}{\Delta_s}\right|_{r=r_c} \ . 
\end{equation}
Since the constraint equations in the Einstein equation give 
the conservation law on the cutoff surface, 
the Brown-York tensor gives the energy-momentum tensor 
to which the solution of the conservation law has already been substituted. 
In order to compare to the stress-energy tensor of fluid on $\mathcal K_n$, 
the Brown-York tensor should be rewritten in terms of the local quantities. 
We use the proper frequency $\omega_c = f^{-1/2}(r) \omega$, 
and take the redshift factor into account as 
\begin{align}
 {f_0}^\text{(proper)} &= \frac{1}{\sqrt{f(r)}} f_0 \ ,&
 {{T^i}_0}^\text{(proper)} &= \frac{r}{\sqrt{f(r)}} {T^i}_0 \ . 
\end{align}
In what follows, we will use these proper quantities for the expressions of the Brown-York tensor. 
Since the Brown-York tensor has a complicated expression, 
we do not write down the full expression, here. 
As we have done for the master field, 
we consider the cut-off surface near the horizon 
and expand the Brown-York tensor in terms of $(r-r_H)$. 
In the  near horizon expansion, 
the sound pole $\Delta_s$ is expressed as 
\begin{equation}
 \Delta_s 
 \sim 
 \left(-\omega _c^2 + \frac{k^2 f'(r)}{2 n r f(r)}\right)
 A(k)
 -\frac{2 i (n-1) \sqrt{f(r)} \left(k^2-K n\right) \omega_c}{n r^2 f'(r)} , \label{BYSoundPole}
\end{equation}
where 
\begin{equation}
 A(k) = 
 1+\frac{2 k^2}{n r f'(r)}
 \sim 1+\frac{k^2}{2\pi n r_H T_H} \ , 
\end{equation}
and we have extracted only the leading order terms of the near horizon expansion 
for each order term in $k$ and $\omega_c$. 
This expression is consistent with the sound pole of fluid \eqref{FluidSoundPole} 
except for the factor of $A(k)$, 
which gives a higher derivative correction. 
The fluid structure generally appears in long wavelength limit of $k\to 0$, 
and the higher derivative correction can usually be neglected in this limit. 
In this sense, \eqref{BYSoundPole} is consistent to the pole of fluids. 
It should be noticed that the angular momentum takes only discrete values for $K>0$. 
In this case, the first order hydrodynamics gives an approximative description 
if $k$ satisfies the condition: 
\begin{equation}
 k^2 < 2\pi n r_H T_H \ . 
\end{equation}
Although we can take the long wavelength limit by taking the radius of 
the surface to be large, 
it is also related to the Hawking temperature. 
In the case of the Schwarzschild black holes, for example, 
$A(k)$ becomes 
\begin{equation}
 A(k) \sim 1 + \frac{2 k^2}{n(n-1)} , 
\end{equation}
and hence the higher derivative terms cannot be neglected for small $n$. 
It would be expected that this additional factor is reproduced 
if the higher derivative terms in fluid mechanics are taken into account. 
We will discuss it in Section~\ref{sec:HigherDeriv}. 

The transport coefficients can be read off 
by comparing \eqref{BYSoundPole} and \eqref{FluidSoundPole}. 
The background energy density and pressure are 
calculated from the background Brown-York tensor \eqref{bgBY}.
In the near horizon expansion, we have 
\begin{equation}
 \bar\varepsilon + \bar P = \frac{f'(r)}{2 \sqrt{f(r)}} + \mathcal O((r-r_H)^{1/2}) \ .
\end{equation}
Then, the speed of sound $c_s$, shear viscosity $\eta$, 
and bulk viscosity $\zeta$ are obtained as 
\begin{align}
 c_s^2 &= \frac{rf'(r)}{2nf(r)} \ , & 
 \eta &= \frac{1}{16\pi G} \ , & 
 \zeta &= 0 \ . \label{Transport}
\end{align}
Here, we wrote the Newton constant $G$, explicitly. 

The numerators $\mathcal N_I$ are calculated as 
\begin{subequations}\label{BYNum}
\begin{align}
 \mathcal N^0_0 
 &\sim 
 \frac{f' k^2}{4 r^2 \sqrt{f}}
 \left(A(k)\right)^2
 f^0_0
 + \frac{i f' \omega_c k}{2r \sqrt{f}}
 \left(A(k)\right)^2
 f_0
\notag\\&\quad
 +\biggl(-\frac{n f' \omega_c^2}{2 \sqrt{f}}A(k) 
  + \frac{k^2 (n-1)\left(k^2-K n\right)}{n r^3 \sqrt{f}} 
 + \frac{i (n-1)\omega_c \left(k^2-K n\right)}{r^2}
 \biggr)
 A(k)
 H_L
\notag\\&\quad
 + 
 \left(\frac{k^2 (n-1)\left(k^2-K n\right)}{n r^3 \sqrt{f}}
 + \frac{i (n-1)\omega_c \left(k^2-K n\right)}{n r^2}\right)
 A(k)
 H_T \ , 
\displaybreak[1]
\\
 \mathcal N_0
 &\sim
 \frac{i f' \omega_c k}{4 r \sqrt f}
 \left(A(k)\right)^2
 f^0_0
 - \frac{f' \omega_c^2}{2 \sqrt f}
 \left(A(k)\right)^2
 f_0
 - \frac{i f^{\prime 2}\omega_c k}{4 f^{3/2}}
 A(k)
 H_L
\notag\\&\quad
 - \left(
 \frac{i (n-1)\omega_c k (k^2 - nK)}{n^2 r^2 \sqrt f}
 + \frac{(n-1)\omega_c^2 (k^2 - n K)}{n r k}
 \right)
 A(k)
 H_T \ , 
\displaybreak[1]
\\
 \mathcal N_L
 &\sim
 - \frac{f^{\prime 2}k^2}{8 n r f^{3/2}}
 A(k)
 f^0_0
 - \frac{i f^{\prime 2}\omega_c k}{4 n f^{3/2}}
 A(k)
 f_0
\notag\\&\quad
 + \left(
 \frac{r f^{\prime 2} \omega_c^2}{4 f^{3/2}}
 - \frac{i (n-1) f' \omega_c (k^2 - nK)}{2 n r f} 
 \right)
 A(k)
 H_L
\notag\\&\quad
 - \frac{i (n-1) f' \omega_c (k^2 - nK)}{2 n^2 r f}
 A(k) 
 H_T \ , 
\displaybreak[1]
\\
 \mathcal N_T
 &\sim
  \left(
 \frac{k^4}{2 n r^3 \sqrt f}
 -\frac{i \omega_c k^2}{2 r^2}
 \right)
 A(k)
 f^0_0
 + 
 \left(
 \frac{i \omega_c k^3}{n r^2 \sqrt f}
 + \frac{\omega_c^2 k}{r}
 \right)
 A(k)
 f_0
\notag\\&\quad
 + \frac{i f' \omega_c k^2}{2 r f}
 A(k)
 H_L
 + \left(-i\omega_c^3 +\frac{i f' \omega_c k^2}{2 n r f}\right)
 A(k)
 H_T \ . 
\end{align}
\end{subequations}
Here, we have written only leading order terms in the near horizon expansion 
for each coefficients of the metric perturbations but 
also included some $\mathcal O((r_c-r_H)^{1/2})$ corrections 
which are relevant to comparison with \eqref{FluidNum}. 
By using the transport coefficients \eqref{Transport}, 
these expressions agree with \eqref{FluidNum} 
except for the factor of $A(k)$ 
and some higher order terms in $k$ and $\omega_c$. 
Therefore, 
up to the higher derivative terms, 
the Brown-York tensor on the cutoff surface near the horizon
is consistent to the energy-momentum tensor of fluid. 
In the next section, we will discuss the higher derivative corrections, 
and then, the extra factor of $A(k)$ and other higher derivative terms 
\eqref{BYNum} are reproduced by the stress-energy tensor of the fluid. 

\section{Higher derivative corrections}\label{sec:HigherDeriv}

In this section, we comment on the higher derivative terms in 
the Brown-York tensor. 
The Brown-York tensor has the structure of the stress-energy tensor 
of fluid in the hydrodynamic regime. 
However, the higher order terms in \eqref{BYSoundPole} and \eqref{BYNum} 
are absent from the linear response of the fluid stress-energy tensor, 
\eqref{FluidSoundPole} and \eqref{FluidNum} because the viscous stress tensor (\ref{ViscousStress}) is composed of only the first order derivative terms. 
Here, we show that the higher derivative terms in the Brown-York tensor can be 
reproduced by adding suitable higher derivative terms to the constituent relation 
(or equivalently to the viscous stress tensor). 

In order to reproduce the sound pole in the Brown-York tensor, 
we introduce an vector field $\tilde u^\mu$ as 
\begin{align}
 u^\mu &= \tilde u^\mu - c_1 \Delta^{\mu\nu}\nabla_\nu\nabla_\rho\tilde u^\rho . 
\end{align}
Then, we keep \eqref{FluidStress}, \eqref{ViscousStress} and \eqref{Projection} unchanged but 
replace the velocity vector $u^\mu$ in \eqref{Shear} and \eqref{Expansion} by $\tilde u^\mu$:  
\begin{subequations}\label{CorrectShearExpan}
\begin{align}
 \tau_{\mu\nu} 
 &= 
 - 2 \eta \tilde\sigma^{\mu\nu} 
 - \zeta \tilde\theta \Delta^{\mu\nu} ,
\\
 \tilde\sigma^{\mu\nu} 
 &= 
 \frac12 \Delta^{\mu\alpha} \Delta^{\nu\beta} 
 \left(
  \nabla_\alpha \tilde u_\beta + \nabla_\beta \tilde u_\alpha 
  - \frac{2}{n} g_{\alpha\beta} \nabla_\gamma \tilde u^\gamma
 \right) , 
\\
 \tilde\theta &= \nabla_\alpha \tilde u^\alpha .  
\end{align}
\end{subequations}
Here, we have defined the projection $\Delta^{\mu\nu}$ with 
the original velocity vector $u^\mu$. 
It should be noticed that the result is not affected 
even if we define $\Delta^{\mu\nu}$ with $\tilde u^\mu$. 
Then, by solving the conservation law, 
we obtain, 
\begin{align}
 \tilde u
 &=
 \frac{1}{\Delta_s}\frac{i\omega}{k} 
 \biggl[-\frac{1}{2}k^2f^0_0 -i\omega k f_0 
 +\frac{2(n-1)}{n}(k^2-nK)i\omega \etahat H_T 
\notag\\&\qquad\qquad\qquad\qquad
 +nk^2\left(c_s^2(1 + c_1 k^2) - c_1 \omega^2 -i\omega \zetahat\right)H_L
 \biggr]\ ,
\\
\delta \varepsilon 
&=\frac{(\ebar+\pbar)(1+c_1 k^2)}{\Delta_s} 
 \biggl[-\frac{1}{2}k^2f^0_0-i\omega k f_0+\frac{2(n-1)}{n}(k^2-nK)i\omega \etahat H_T
\notag\\&\qquad\qquad\qquad\qquad\qquad
 +n\left(\frac{2(n-1)}{n}(k^2-nK)i\omega \etahat +\omega^2\right)H_L\biggr] \ ,
\\
 \Delta_s &= \left(-\omega^2+c_s^2 k^2\right)(1 + c_1 k^2)
 -i\omega \left(\Gamma_s k^2 -2(n-1) K\etahat\right) \ . 
\end{align}
By taking 
\begin{equation}
 c_1 = \frac{2}{n r f'(r)}\ ,
\end{equation}
the pole in the Brown-York tensor \eqref{BYSoundPole} is reproduced. 
Since $\tilde u^\mu$ can be expressed in terms of $u^\mu$ 
as 
\begin{align}
 \tilde u^\mu &= \left(1-c_1 \Delta^{\mu\nu}\nabla_\nu \nabla_\rho\right)^{-1} u^\rho 
\notag\\ &= u^\mu + \sum_{n=0}^{\infty} c_1^{n+1} \Delta^{\mu\nu}\nabla_\nu 
 \left(\nabla_\alpha \Delta^{\alpha\beta} \nabla_\beta\right)^n \nabla_\rho u^\rho \ , 
\end{align}
\eqref{CorrectShearExpan} can be interpreted as 
the shear and expansion with higher derivative corrections. 

In order to reproduce the higher order corrections in 
the numerators of the Brown-York tensor \eqref{BYNum}, 
we further introduce correction terms which is related to the curvature. 
For the Riemann tensor,
 two indices must be contracted by 
projection tensor $\Delta^{\mu\nu}$, or equivalently, $u^\mu$. 
Since the correction terms appear only in the viscous stress, 
the other indices must be projected by $\Delta^{\mu\nu}$.
Then the possible correction terms are expressed in terms of 
$u^\rho u_\sigma {R_{i\rho j}}^\sigma$ or $R_{ij}$.%
\footnote{%
Similar correction terms are considered in the framework of 
the relativistic hydrodynamics and its relation to the AdS/CFT correspondence \cite{Baier:2007ix}. 
}
In the linear response theory, we have 
\begin{align}
 \delta \mathcal R_{i}^j 
 &= 
 - \delta (u^\rho u_\sigma {R_{i \rho}}^{j \sigma})
\notag\\
 &= 
 \left(
  - \frac{k^2}{2} f^0_0 - i \omega k f_0 - \omega^2 n H_T
 \right) \Sbb_{i}^j 
 + \frac1n 
 \left(
  \frac{k^2}{2} f^0_0 + i \omega k f_0 - \omega^2 n H_L
 \right) \delta^j_i \Sbb \ , 
\\
 \delta {R^i}_{j} - \delta {\mathcal R^i}_{j} 
 &= 
 \delta {R_{il}}^{jl} 
\notag\\
 &= 
 - (n-2) k^2 \left(
  H_L + \frac1n H_T 
 \right) \Sbb_{i}^j
\notag\\
&\qquad
 + 2 \frac{n-1}{n} (k^2-nK)
 \left(
  H_L + \frac1n H_T
 \right) \delta_i^j \Sbb \ .  
\end{align}
Then the viscous stress tensor is 
\begin{equation}
 \tau^{\mu\nu} 
 = 
 - 2 \eta \tilde \sigma^{\mu\nu} 
 - \zeta \tilde \theta \Delta^{\mu\nu} 
 + c_2 \mathcal R_L \Delta^{\mu\nu} + c_3 \mathcal R_T^{\mu\nu} 
 + c_4 R_L \Delta^{\mu\nu} + c_5 R_T^{\mu\nu}  
\end{equation}
where 
\begin{align}
 \mathcal R_L &= \frac{1}{n} \delta{\mathcal R_i}^i
\\
 \mathcal R_T^{\mu\nu} 
 &= 
 \Delta^{\mu\alpha} \Delta^{\nu\beta} 
 \left(
  \delta\mathcal R_{\alpha \beta} - g_{\alpha \beta}\mathcal R_L
 \right) , 
\\
 R_L &= \frac{1}{n} \delta {R_i}^i
\\
 R_T^{\mu\nu} 
 &= 
 \Delta^{\mu\alpha} \Delta^{\nu\beta} 
 \left(
  \delta R_{\alpha \beta} - g_{\alpha \beta} R_L
 \right) .  
\end{align}
Then, by taking the coefficients $c_2$, $c_3$, $c_4$ and $c_5$ as 
\begin{align}
 c_2 &= \frac{r}{\sqrt f} \ , &
 c_3 &= - \frac{2 \sqrt f}{f'} \ , &
 c_4 &= \frac{r}{2 \sqrt f} \ , & 
 c_5 &= - \frac{\sqrt f}{f'} \ , 
\end{align}
the stress-energy tensor agrees with the Brown-York tensor 
for the leading terms in the near horizon expansion \eqref{BYNum} 
but including the higher order terms of $k$ and $\omega$.

\section{Counter terms and junction condition}\label{sec:CounterTerm}

In the preceding sections, we have seen that 
the Brown-York tensor on the cutoff surface near the horizon 
provides the stress-energy tensor of fluid. 
In order to define the energy-momentum tensor on the surface 
by using the Brown-York tensor, 
we need to introduce additional counter terms on the surface. 
These terms are called as the counter terms 
in the framework of the holographic renormalization. 
These boundary terms corresponds to taking an appropriate boundary condition, 
and hence they are equivalent to fixing the reference geometry. 
In the framework of the membrane paradigm, 
the energy-momentum tensor is defined by the Israel junction condition and 
is given by the difference of the Brown-York tensors 
which are calculated in each sides of the surface. 
The energy-momentum tensor of the fluid comes from 
the Brown-York tensor in the black hole geometry. 
The effects of the geometry inside the surface correspond 
to the counter terms in the holographic renormalization. 

The counter terms, or equivalently, contributions from 
the geometry inside the stretched horizon, 
give additional contributions to the fluid. 
We do not consider the details of this effect, 
but nonetheless make a few comments on it in this section. 

In the case of the AdS/CFT correspondence, 
the counter terms are determined by assuming that 
they are local covariant functions of the intrinsic geometry. 
The most simple counter term for the action is the cosmological constant, 
\begin{equation}
 S_\text{ct} = C_\text{ct} \int d^{n+1} x \sqrt{-\gamma} , \label{CTcc}
\end{equation}
where $C$ is a coefficient which depends on $r_c$. 
This counter term gives an additional term to the energy-momentum tensor: 
\begin{equation}
 T^{\mu\nu}_\text{ct} = C_\text{ct} \gamma^{\mu\nu} 
\end{equation}
and the coefficient is fixed by requiring cancellation of divergences. 
This term gives corrections to the energy density and pressure as 
\begin{align}
 \delta \varepsilon &= -C_\text{ct} \ , & 
 \delta P &= C_\text{ct} . 
\end{align}
The linear response of the stress-energy tensor 
depends only on the combination of $(\bar\varepsilon + \bar P)$. 
Therefore, the counter term \eqref{CTcc} does not 
affect the fluid on the surface. 

Another simple counter term is the intrinsic curvature on the surface. 
This term gives only the higher derivative terms, 
and hence, does not contribute to the fluid in the first order formalism, 
which we have mostly discussed in this paper. 
For finite $r_c$, we can also modify the fluid theory by adding the curvature term. 

The pole structure in the Brown-York tensor 
comes from the term proportional to the master field. 
Since the master field represents the propagating mode in the bulk, 
the counter terms do not affect the pole structure as long as 
they are defined by local functions on the intrinsic geometry on the surface. 
Therefore, the counter terms basically are not important for the fluid. 

In the membrane paradigm, 
there are additional contributions from the geometry inside the stretched horizon. 
The inside geometry is given by the geometry with $M=0$. 
For example, in the case of $\Lambda=0$, the metric is 
\begin{subequations}\label{inside}
\begin{align}
 ds^2 
 &= 
 - d\tilde t^2 + d\tilde r^2 + \tilde r^2 d\tilde\sigma_n^2 \ , 
\\
 d\tilde\sigma_n^2 
 &= 
 \hat\gamma_{ij} d\tilde z^i d\tilde z^j
\end{align}
\end{subequations}
The relation between the coordinates inside and outside the surface 
is determined by requiring that the induced metrics calculated in each side agrees. 
For the background metric, 
we obtain 
\begin{align}
 \tilde t &= \sqrt{f(r_c)}\, t \ , & 
 \tilde r &= r \ , & 
 \tilde z^i &= z^i \ . 
\end{align}
The energy-momentum tensor on the surface is obtained 
by subtracting the Brown-York tensor in the inside geometry 
from that in the outside geometry. 
Then, the corrections to the energy density and pressure are 
\begin{align}
 \delta \bar\varepsilon &= \frac{n}{r} \ , & 
 \delta \bar P &= - \frac{n(n-1)}{r} \ . 
 \label{junction}
\end{align}
The linear response of the stress-energy tensor 
depends only on the combination of $(\bar\varepsilon + \bar P)\sim \mathcal O((r_c-r_H)^{-1/2})$. 
Therefore, the contributions from the inside geometry does not contribute 
to the leading order terms of the near horizon expansion. 
It should also noticed that this correction is necessary to obtain positive energy density 
although it is not important for the linear responses. 

If we introduce the perturbations of the metric directly to \eqref{inside} with \eqref{junction}, 
the contributions from the inside geometry give 
similar contributions to those from the outside geometry. 
We can calculate the Brown-York tensor in the same fashion to that in the black hole geometry, 
but the ingoing boundary condition at the horizon 
should be replaced by the regularity condition at $r=0$. 
The Brown-York tensor in inside geometry is calculated separately from that in outside geometry, 
and does not affect the fluid stress-energy tensor, but gives simply additional terms. 
These terms do not have pole at $k\to 0$ and $\omega\to 0$ and do not give 
additional poles of a fluid at least for $K\neq 0$. 
They can be interpreted as higher order corrections of the derivative expansion. 
This effect appears because the inside geometry is not empty but 
there is an propagating mode. 

In order to take the geometry inside the surface to be completely flat, 
the effect of the metric perturbation should be absorbed into the junction condition. 
By taking the coordinates inside the surface as 
\begin{align}
 \tilde t &= \sqrt{f(r_c)}\, t + Y^0 e^{-i\omega t}\,\Sbb \ ,\\
 \tilde r &= r + Y^r e^{-i\omega t}\,\Sbb \ ,\\
 \tilde z &= z^i + \frac{1}{r} L e^{-i\omega t}\,\Sbb^i \ ,
\end{align}
the metric of the flat space \eqref{inside} is expressed as 
\begin{align}
 ds^2 
 &= 
 - f(r_c)d t^2 + d r^2 + r^2 d \sigma_n^2 + h_{\mu\nu}dx^\mu dx^\nu \ , \label{InsideFlat}
\end{align}
where the perturbations are 
\begin{align}
 h_{00} &= 2i\omega Y_0 e^{-i\omega t}\,\Sbb \ ,\\
 h_{0i} &= (i \omega r L + k Y_0) e^{-i\omega t}\,\Sbb_i \ ,\\
 h_{ij} 
 &= 
 2 r e^{-i\omega t}\left[
 \left(-\frac{k}{n} L - Y_r\right)\hat\gamma_{ij}\Sbb 
 + k L \Sbb_{ij}
 \right]  \ . 
\end{align}
Although we have taken the gauge $h_{r\mu} = 0$ on the outside geometry, 
the gauge condition on the inside geometry can be different, and in fact 
$h_{r\mu}$ is non-zero in \eqref{InsideFlat}. 
Since these coordinates have only three free parameters, 
it cannot reproduce arbitrary metric perturbations. 
However, they can provide non-vanishing master field $\Phi$ in the outside geometry, 
and the energy-momentum tensor has the structure of fluid. 
The corrections from the inside geometry do not have any pole, 
and can be treated in a similar fashion to the counter terms.

\section{Conclusion and discussions}\label{sec:conclusion}

In this paper, we have studied the correspondence between 
black holes and fluids. 
Our analysis is based on the Fluid/Gravity correspondence, 
but can also be interpreted as the membrane paradigm. 
Since the bulk viscosity cannot be calculated in the Rindler limit 
due to the incompressibility, 
we have put the cutoff surface at a finite distance from the horizon. 
Then, the geometries are no longer universal, and hence, 
we have focused on simplest cases. 
We have considered maximally symmetric black holes, 
which contain asymptotically non-AdS geometries with compact horizons. 
For these black holes, it is expected that the correspondence 
appears only near the horizon. 
In order to avoid the incompressibility, 
we have put the cutoff surface near the horizon but 
have kept a finite distance from the horizon, 
and then, considered the near horizon expansion. 
Generally, the leading order terms of the near horizon expansion 
gives the Rindler limit. 
However, due to the incompressibility, 
the leading order contributions give 
the corrections to the Rindler limit for the sound modes. 
We have considered the correspondence of the sound modes 
in the framework of the linear response theory. 
The energy-momentum tensor on the surface 
is given by the Brown-York tensor in both frameworks of 
the Fluid/Gravity correspondence and membrane paradigm. 
Since the Einstein equation contains the constraint equations  
which correspond to the conservation law of 
the energy-momentum tensor on the surface, 
the equation of continuity and the linearized Navier-Stokes equation are 
imposed on the stress-tensor of the fluid. 
Then, we have shown that the linear responses 
of the Brown-York tensor and fluid stress-energy tensor agree, 
at the leading order of near horizon expansion. 
Since these contributions contain the corrections to the Rindler limit, 
the fluid becomes compressible. 
There are additional contributions from 
the counter terms in the Fluid/Gravity correspondence 
or the Brown-York tensor on the geometry inside the stretched horizon 
in the membrane paradigm. 
These contributions give additional terms 
but do not modify the pole structure of 
the Brown-York tensor in the black hole geometry. 
Therefore, the counter terms or the effects of the inside geometry 
do not modify the fluid structure on the surface. 

We have calculated the transport coefficients in the near horizon expansion. 
At the leading order, the bulk viscosity vanishes: $\zeta = 0$. 
This result is different from the previous result in the membrane paradigm. 
In the membrane paradigm, the energy-momentum tensor 
on the stretched horizon was calculated and the limit in which 
the stretched horizon becomes the true horizon was taken. 
In this limit, the surface becomes null and the normal vector becomes 
the only vector which is pointing the causal direction on the surface. 
Then, in the previous studies of the membrane paradigm, 
the normal vector $n^\mu$ to the horizon was identified to the velocity vector of fluid. 
Since the stress-energy tensor is given by \eqref{eq:adsby}, 
the energy $\varepsilon$, pressure $P$, viscous stress $\tau^{\mu\nu}$, 
shear $\sigma^{\mu\nu}$, and expansion $\theta$ were identified as 
\begin{subequations}\label{OldMembrane}
\begin{align}
 \varepsilon 
 &= 
 \theta \ , 
\\
 P 
 &= 
 - u^\mu u^\nu \mathcal K_{\mu\nu} \ , 
\\
 \tau^{\mu\nu} 
 &= - \sigma^{\mu\nu} + \frac{n-1}{n} \Delta^{\mu\nu} \theta \ , 
\\
 \sigma^{\mu\nu}
 &=
 \Delta^{\mu\rho}\Delta^{\nu\sigma}\mathcal K_{\rho\sigma} \ , 
\\
 \theta 
 &= 
 \Delta^{\rho\sigma} \mathcal K_{\rho\sigma} \ .  
\end{align}
\end{subequations}
However, this identification is correct only when 
the stretched horizon becomes the true horizon. 
Since the stretched horizon is defined as an timelike surface, 
the normal vector is not tangent to it. 
Then, the above identification must be modified. 
For the shear modes, 
\eqref{OldMembrane} gives the correct result at the leading order of 
the near horizon expansion. 
For the sound modes, however, 
the fluid becomes incompressible as has been discussed 
in the Fluid/Gravity correspondence, 
and hence, the next-to-leading order corrections become important. 
In fact, the leading order contributions in this paper 
should be interpreted as the next-to-leading order correction 
to the vanishing leading terms in comparison with the shear modes. 
Therefore, we cannot use the above identification for the sound modes, 
and hence, obtained the different result, $\zeta=0$. 

It should also be noticed that there is an ambiguity in 
the identification of the pressure and extension, in \eqref{OldMembrane}. 
Since both of them come from the trace part of 
the spatial components of the stress tensor, 
they cannot easily be distinguished. 
For the sound modes of fluid, the speed of sound is 
calculated from the ratio of the variations of 
the energy density and pressure, and expected to be constant. 
This implies that the pressure also has a term proportional to 
the expansion since the energy density equals to the expansion in \eqref{OldMembrane}. 
Then, the coefficient of the expansion in $\tau^{\mu\nu}$ in \eqref{OldMembrane} 
should not simply be identified to the bulk viscosity 
but might contain a contribution which should be identified as a part of the pressure. 
This fact gives additional correction to the negative bulk viscosity. 
By taking into account the difference between the normal vector and velocity vector 
on the timelike stretched horizon, 
the energy density is no longer proportional to the expansion. 
Then, in our analysis, the pressure and bulk viscosity can be uniquely identified. 

It is interesting that the bulk viscosity vanishes even for 
the asymptotically non-AdS geometries. 
In the case of AdS, it could be a consequence of the conformal symmetry. 
However, in this paper, we have considered the geometries 
which are not asymptotically AdS. 
Then, the zero bulk viscosity does not come from the conformal symmetry. 
It is natural to expect that the bulk viscosity vanishes 
in more general geometries. 
Since in order to calculate the bulk viscosity, 
the cutoff surface must be at a finite distance from the horizon, 
the fluid cannot be universal in these cases. 
It is interesting to consider the case of other geometries. 
It should be noticed that the non-zero bulk viscosity can easily be obtained 
by considering the dimensional reduction of the zero bulk viscosity cases. 
This implies that the bulk viscosity becomes non-zero if 
there is a dilation in the gravity side. 

In this paper, we have considered only the linear responses of the fluid. 
There remains a possibility that there are other modes 
which give the negative bulk viscosity. 
However, the sound modes which has been analyzed in this paper 
are present independently from such modes. 
We have also discussed possible higher derivative corrections 
for the fluid stress-energy tensor. 
It would be also interesting how they can be understood 
in the framework of the relativistic fluid \cite{Israel:1976tn, Israel:1979wp}%
\footnote{%
Higher derivative corrections in the AdS/CFT correspondence are discussed in 
\cite{Bhattacharyya:2008jc, Baier:2007ix, Natsuume:2007ty, Kuperstein:2013hqa}.
}. 
It is also straightforward to calculate the higher order corrections 
in the near horizon expansion. 
Although the higher order corrections in $k$ appear in this paper, 
there are no higher order corrections in $\omega$. 
They appear with factor of $f$ and can be calculated 
by considering the higher order corrections of the near horizon expansion.  
For example, the relaxation time which appears as a coefficient of 
$u^\rho D_\rho \sigma^{\mu\nu}$ will appear in the higher order corrections. 
These are left for future studies. 

\subsubsection*{Acknowledgments}

The work of Y.M. is supported by JSPS Research Fellowship for Young Scientists and 
in part by Grant-in-Aid for JSPS Fellows (No.23-2195).

\appendix

\section{Full expressions}\label{apdx:fullex}

By using the Einstein equation and the definition of the master field, 
the first derivatives of the metric perturbations can be expressed 
in terms of the metric perturbations and master field without derivatives. 
Here, we define $\hat f_0 = f_0/k$ and $\hat H_T = H_T/k^2$. 
The full expression of the relation is the following: 
\begin{align}
 f^0_0{}'
 &= 
 \frac{k^2}{n r f(r)} f^0_0
 + \frac{2 i k^2 \omega }{n f(r)^{3/2}} \hat f_0
\notag\\&\qquad
  + \biggl\{\frac{(n-1) f(r) \left(k^2-2 K n\right)-k^2 r f'(r)}{n r f(r)^2}
\notag\\&\qquad\qquad\qquad
 -\frac{r^2 f'(r)^2-2
   r f(r) f'(r)-\left(n^2-1\right) f(r)^2}{2 r f(r)^2}-\frac{2 r \omega ^2}{f(r)}\biggr\}
 H_L
\notag\\&\qquad
 + \biggl\{
 \frac{k^2 \left((n-1) f(r) \left(k^2-2 K n\right)-k^2 r f'(r)\right)}{n^2 r
   f(r)^2}
\notag\\&\qquad\qquad\qquad
 -\frac{k^2 \left(r^2 f'(r)^2-2 r f(r) f'(r)-\left(n^2-1\right)
   f(r)^2\right)}{2 n r f(r)^2}
 \biggr\}\hat H_T
\notag\\&\qquad
 + \biggl\{\frac{k^2 r^{-\frac{n}{2}-1} \left(-r f'(r)-n f(r)+f(r)\right) \left((n+1) f(r)-r
   f'(r)\right)}{4 n f(r)^2}
\notag\\&\qquad\qquad\qquad
 -\frac{k^4 r^{-\frac{n}{2}-1} \left(-r f'(r)-n
   f(r)+f(r)\right)}{2 n^2 f(r)^2} \biggr\} \Phi \ ,
\displaybreak[1]
\\
 f_0'
 &= 
 -\frac{n f(r)-r f'(r)}{r f(r)} \hat f_0
 -\frac{2 i \omega }{\sqrt{f(r)}} H_L
\notag\\&\qquad
 +\biggl\{ \frac{i \omega  \left((n+1) f(r)-r f'(r)\right)}{\sqrt{f(r)}}-\frac{2 i k^2 \omega
   }{n \sqrt{f(r)}} \biggr\} \hat H_T
\notag\\&\qquad
 + \biggl\{\frac{i k^2 \omega  r^{-n/2}}{n \sqrt{f(r)}}-\frac{i \omega  r^{-n/2} \left((n+1)
   f(r)-r f'(r)\right)}{2 \sqrt{f(r)}} \biggr\} \Phi \ ,
\displaybreak[1]
\\
 H_L'
 &= 
 \biggl\{ \frac{k^2}{n r f(r)}-\frac{-r f'(r)+n f(r)+f(r)}{2 r f(r)}\biggr\} H_L
\notag\\&\qquad
 + \biggl\{ \frac{k^4}{n^2 r f(r)}-\frac{k^2 \left((n+1) f(r)-r f'(r)\right)}{2 n r f(r)}\biggr\} \hat H_T
\notag\\&\qquad
 + \biggl\{\frac{k^2 r^{-\frac{n}{2}-1} \left((n+1) f(r)-r f'(r)\right)}{4 n f(r)}-\frac{k^4
   r^{-\frac{n}{2}-1}}{2 n^2 f(r)} \biggr\} \Phi \ ,
\displaybreak[1]
\\
 H_T'
 &= 
 \biggl\{\frac{k^2}{2 (n-1) r f(r) \left(K n-k^2\right)}-\frac{n \left((n+1) f(r)-r
   f'(r)\right)}{4 (n-1) r f(r) \left(K n-k^2\right)}\biggr\} f^0_0
\notag\\&\qquad
 + \biggl\{\frac{i k^2 \omega }{(n-1) f(r)^{3/2} \left(K n-k^2\right)}-\frac{i n \omega 
   \left((n+1) f(r)-r f'(r)\right)}{2 (n-1) f(r)^{3/2} \left(K n-k^2\right)} \biggr\} \hat f_0
\notag\\&\qquad
 + \biggl\{\frac{(n-1) f(r) \left(3 k^2-2 K n\right)-k^2 r f'(r)}{2 (n-1) r f(r)^2 \left(K
   n-k^2\right)}
\notag\\&\qquad\qquad\qquad
 -\frac{n \left(-2 n r f(r) f'(r)+r^2 f'(r)^2+\left(n^2-1\right)
   f(r)^2\right)}{4 (n-1) r f(r)^2 \left(K n-k^2\right)} \biggr\} H_L
\notag\\&\qquad
 + \biggl\{-\frac{k^2 \left(-2 n r f(r) f'(r)+r^2 f'(r)^2+\left(n^2-1\right) f(r)^2\right)}{4
   (n-1) r f(r)^2 \left(K n-k^2\right)}
 -\frac{n r \omega ^2 \left((n+1) f(r)-r
   f'(r)\right)}{2 (n-1) f(r) \left(K n-k^2\right)}
\notag\\&\qquad\qquad\qquad
 +\frac{k^2 \left((n-1) f(r)
   \left(3 k^2-2 K n\right)-k^2 r f'(r)\right)}{2 (n-1) n r f(r)^2 \left(K
   n-k^2\right)}+\frac{k^2 r \omega ^2}{(n-1) f(r) \left(K n-k^2\right)} \biggr\} \hat H_T
\notag\\&\qquad
 + \biggl\{ \frac{n \omega ^2 r^{1-\frac{n}{2}} \left((n+1) f(r)-r f'(r)\right)}{4 (n-1) f(r)
   \left(K n-k^2\right)}
\notag\\&\qquad\qquad\qquad
-\frac{k^2 r^{-\frac{n}{2}-1} \left((n-1) f(r)-r
   f'(r)\right) \left((n+1) f(r)-r f'(r)\right)}{8 (n-1) f(r)^2 \left(k^2-K
   n\right)}
\notag\\&\qquad\qquad\qquad
 +\frac{k^4 r^{-\frac{n}{2}-1} \left((n-1) f(r)-r f'(r)\right)}{4 (n-1) n
   f(r)^2 \left(k^2-K n\right)}
 +\frac{k^2 \omega ^2 r^{1-\frac{n}{2}}}{2 (n-1) f(r)
   \left(k^2-K n\right)}\biggr\} \Phi \ . 
\end{align}

\end{document}